\documentclass[sigconf]{acmart}
\AtBeginDocument{%
  }

\usepackage{graphicx} 
\graphicspath{{figures/}} 
\usepackage{multirow}
\usepackage{dsfont}
\usepackage{colortbl}
\usepackage[ruled,linesnumbered]{algorithm2e}
\usepackage{booktabs}

\usepackage{float}

\usepackage{caption}
\usepackage{subcaption}

\usepackage{balance}

\usepackage{cleveref}
\crefname{section}{\S}{\S\S}
\crefname{table}{Table}{}
\crefname{figure}{Figure}{}
\crefname{algorithm}{Algorithm}{}
\crefname{equation}{Eq.}{}
\crefname{appendix}{App.}{}
\crefname{prop}{Proposition}{}
\crefname{thm}{Theorem}{}
\crefformat{section}{\S#2#1#3}  

\copyrightyear{2025}
\acmYear{2025}
\setcopyright{acmlicensed}\acmConference[WSDM '25]{Proceedings of the Eighteenth ACM International Conference on Web Search and Data Mining}{March 10--14, 2025}{Hannover, Germany}
\acmBooktitle{Proceedings of the Eighteenth ACM International Conference on Web Search and Data Mining (WSDM '25), March 10--14, 2025, Hannover, Germany}
\acmDOI{10.1145/3701551.3703563}
\acmISBN{979-8-4007-1329-3/25/03}

\begin{document}

\title{Improving FIM Code Completions via Context \& Curriculum Based Learning}

\author{Hitesh Sagtani}
\orcid{0009-0003-6995-1912}
\affiliation{%
  \institution{Sourcegraph Inc}
  \city{Bengaluru}
  \country{India}
}
\email{sagtanih@gmail.com}

\author{Rishabh Mehrotra}
\authornote{Work done while at Sourcegraph.}
\orcid{0000-0002-0836-4605}
\affiliation{%
  \institution{Pavo AI}
  \city{London}
  \country{UK}
}
\email{erishabh@gmail.com}

\author{Beyang Liu}
\orcid{0009-0000-3325-6465}
\affiliation{%
  \institution{Sourcegraph Inc}
  \city{San Francisco}
  \country{USA}
}
\email{beyang@sourcegraph.com}

\begin{abstract}

Fill-in-the-Middle (FIM) models play a vital role in code completion tasks, leveraging both prefix and suffix context to provide more accurate and contextually relevant suggestions. This paper presents approaches to improve FIM code completion while addressing the challenge of maintaining low latency for real-time coding assistance.
We enhance FIM code completion by incorporating context and curriculum examples in the training process. We identify patterns where completion suggestions fail more frequently, revealing complexities that smaller language models struggle with. To address these challenges, we develop a curriculum dataset by extracting hard-to-complete patterns from code repositories and generate context examples using semantic and static analysis tools (e.g. TSC compiler). We fine-tune various sized models, including StarCoder and DeepSeek, on this enhanced dataset.
Our evaluation encompasses three key dimensions: the Santa Coder FIM task, the Amazon CCEval benchmark, and a new Multi-Line Infilling evaluation benchmark derived from SWE-bench. Comprehensive ablation studies across multiple model sizes reveal that while all fine-tuned models show improvements, the performance gains are more pronounced for smaller parameter models and incorporating \textit{difficult-to-complete} examples, as part of curriculum learning, improves the code completion performance. This finding is particularly significant given the latency constraints of code completion tasks. While larger models like GPT and Claude perform well in multi-line completions but are prohibitively challenging to use given high latency, and our fine-tuned models achieve a balance between performance and latency. Finally, we validate our approach through online A/B testing, demonstrating tangible improvements in Completion Acceptance Rate (CAR) and Completion Persistence Rate (CPR), with zero latency impact.

\end{abstract}



\maketitle

\begin{figure*}[t]
    \centering
     \vspace{-2ex}
    \includegraphics[width=\linewidth]{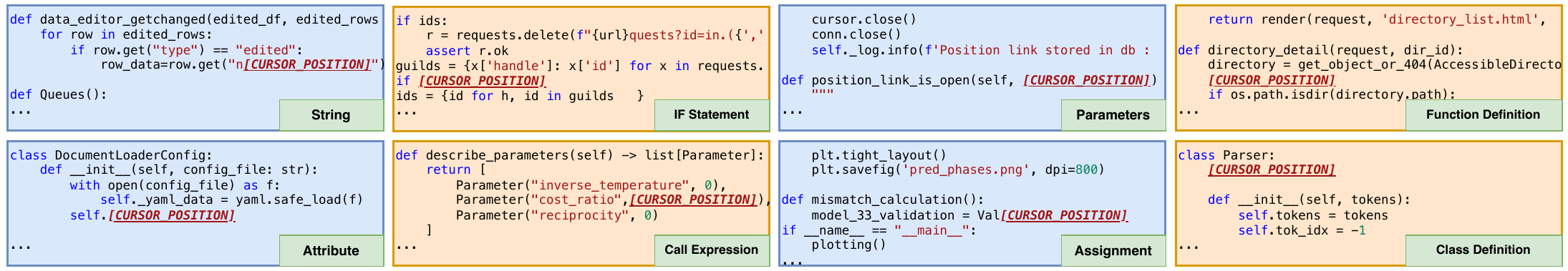}\
    \caption{Illustrative examples of various AST node types. The cursor position highlights the position where completions are triggered, with the node type indicated at the bottom right of each example.}
    \label{fig:ast_diagram}
    \vspace{-1em}
\end{figure*}

\section{Introduction}

Code completion assistants such as Github Copilot, Cody, Cursor etc. have become an integral part of modern software development, significantly enhancing programmer productivity. These assistants, powered by Large Language Models (LLMs), offer a range of features including autocomplete (AC), code editing, unit test generation, and interactive chat capabilities. Among these, autocomplete stands out as one of the most frequently used features by volume. Unlike code editing and unit test generation, autocomplete faces a critical trade-off between latency and model size. The goal is to provide accurate code completions within a stringent 500ms latency constraint, otherwise, users may type the next character, rendering the previous suggestion redundant. The task involves Fill-in-the-Middle (FIM) completion, which differs from traditional left-to-right completion by requiring consideration of both the prefix and suffix around the cursor position. This complexity is compounded by the challenge of evaluating completion success. It necessitates a nuanced understanding of where models excel, where they fall short, and what types of training data are most beneficial.

In our study, we address the challenge of enhancing code completion models within real-world product constraints. We motivate the need for incorporating more challenging and context-aware examples in the Fill-in-the-Middle (FIM) training regime. Our analysis reveals that completion suggestions often fail in cases involving complex code structures. Furthermore, we demonstrate that smaller code language models struggle to provide accurate completions even with optimal context in the prompt. To address these challenges, we introduce a training regime by extracting more complex code patterns from the code repositories using tree-sitter \footnote{\url{https://tree-sitter.github.io/tree-sitter/}} for parsing the code. Additionally, we add context training examples by employing retrieval methods such as BM25. To retrieve precise definitions of code symbols, particularly for TypeScript and JavaScript, we leverage the TypeScript Compiler API\footnote{\url{https://github.com/microsoft/TypeScript/wiki/Using-the-Compiler-API}}. Our findings indicate that incorporating such \emph{difficult-to-complete} examples (i.e. curriculum learning) along with the relevant context, from the repository during training significantly improves code completion quality, especially for smaller models crucial in real-time applications. We apply these methods to fine-tune popular open-source models like StarCoder and DeepSeek, demonstrating the effectiveness of our approach across different model architectures and parameters.

We conduct comprehensive offline and online evaluations of our approach. For offline assessment, we utilize existing datasets like Single-Line Infilling ~\cite{allal2023santacoder} and Amazon's CrossCodeEval (\emph{CCEval}) ~\cite{ding2023crosscodeevaldiversemultilingualbenchmark}, while also introducing a new Multi-Line Infilling dataset derived from SWE-Bench ~\cite{jimenez2023swe}. This new dataset addresses limitations in current benchmarks, which often focus on single-line completions or lack contextual consideration. 

Finally, we present insights from A/B experiments into the behavior of fine-tuned models. We observe directional alignment
between our off- and online experiments, and we show significant improvement in online metrics such as Completion Acceptance Rate (CAR) and Completion Persistence Rate (CPR), while maintaining latency.

In summary, the contributions of our work include the following:

\begin{enumerate}
    \item We characterize FIM code completion problem and motivate the challenges using both synthetic and real-world data (\S\ref{sec:concept}).

    \item We propose a contextual, curriculum-based training regime to enhance pre-trained model performance, particularly benefiting smaller sized models (\S\ref{sec:method}).

    \item We introduce a new Multi-Line Infilling dataset, representing real-world repositories (\S\ref{sec:multline_dataset}).

    \item We present results from live experiments that highlight the effectiveness of our approach, leading to online improvements (\S\ref{sec:applications}).

\end{enumerate}

\vspace{-1em}
\section{Related Work}

\emph{Language Models for Code Generation:} Language models have become integral to modern software development, with both non-code-focused ~\cite{achiam2023gpt, dubey2024llama, jiang2024mixtral} and code-specific models ~\cite{nijkamp2022codegen, wang2021codet5} demonstrating exceptional code generation capabilities. Initially trained on next-token prediction, subsequent research recognized the importance of the Fill-in-the-Middle (FIM) property during pre-training for code completions ~\cite{bavarian2022efficient}. In addition, given that code completion is a latency-sensitive feature, smaller variants (approximately 1 to 16 billion parameters) of models like CodeLlama ~\cite{roziere2023code}, Starcoder ~\cite{li2023starcoder, lozhkov2024starcoder}, and others ~\cite{fried2022incoder, bai2023qwen, allal2023santacoder, guo2024deepseek, zhu2024deepseek, pinnaparaju2024stable, team2024codegemma} have been developed, incorporating the FIM objective into their pre-training loss functions. Recently, the Mixture of Experts (MoE) architecture ~\cite{masoudnia2014mixture} has gained popularity for its ability to reduce latency by having fewer active parameters during inference, as evidenced in models like Codestral ~\cite{codestral2024mistral} and DeepSeek-Coder-v2 ~\cite{zhu2024deepseek}.

\emph{Context Aware Code Completion:} Numerous studies have explored fine-tuning code completion models ~\cite{shi2023towards, guo2024ft2ra, miksik2023fine, mao2024context, nguyen2024information, nandanwar2023contextual} and utilizing RAG pipelines to generate accurate completions.  Notable examples include RepoCoder ~\cite{zhang2023repocoder},  RepoFusion ~\cite{shrivastava2023repofusion} including others ~\cite{hartman2024ai, phan2024repohyper, wu2024repoformer, liu2024graphcoder}. Our work investigates fine-tuning on challenging examples with relevant repo context. Despite various code completion studies from companies like Meta ~\cite{murali2024ai, dunay2024multi, maddila2024ai} and Google ~\cite{google2022ml}, insights from real-world applications remain limited, with most data coming from internal company users. To the best of our knowledge, our work provides the first insights from real-world A/B testing on code completions.

\emph{Code Evaluation Benchmarks:} Recent advancements in code evaluation benchmarks have been crucial for assessing the performance of language models in code generation tasks. HumanEval ~\cite{chen2021evaluating} and MBPP ~\cite{austin2021program} focus on left-to-right completions and evaluate using the \emph{Pass@1} ~\cite{chen2021evaluating} metric. Allal \emph{et al.} ~\cite{allal2023santacoderdontreachstars} introduced the Single-Line infilling benchmark, significant for evaluating code inserts by incorporating suffix information. However, these approaches often fall short of representing real-world scenarios as they lack repo-level code generation capabilities. In response, more recent benchmarks like CrossCodeEval (CCEval) ~\cite{ding2023crosscodeevaldiversemultilingualbenchmark} and others ~\cite{liu2023repobenchbenchmarkingrepositorylevelcode, yu2024codereval} have emerged, assessing repository-level code generation capabilities. They typically use metrics such as \emph{Exact Match}, \emph{Edit Similarity}, \emph{CodeBLEU} ~\cite{ren2020codebleu, post2018call} etc. Despite these advances, these benchmarks are still constrained by their reliance on single-line ground truth evaluations. To address this limitation, we present a Multi-line infilling evaluation benchmark for code generation models sourced from real-world repository.

\section{Motivating analysis} \label{sec:concept}

In this section, we provide a motivating analysis to leverage curriculum based examples and its relevant context while training model for fill-in-the-middle tasks in code completions. We highlight the cases where LLM fails to correctly predict the code more often than the other cases.

\vspace{-1ex}
\subsection{Data Context}
Figure \ref{fig:ast_diagram} shows examples of various code suggestion scenarios. The Cursor Position indicates where a completion request is triggered, with preceding code as the \emph{Prefix} and subsequent code as the \emph{Suffix}. We also note that, the code in a file can be parsed into an Abstract Syntax Tree (AST). We highlight nodes of AST in figure \ref{fig:ast_diagram} for each example in the green box.

Our synthetic dataset for analysis consists of tuples (\emph{P}, \emph{S}, \emph{M}, \emph{C}), where \emph{P} is the \emph{Prefix}, \emph{S} is the \emph{Suffix}, \emph{M} is the ground truth completion, and \emph{C} is the context (relevant files from the same repository). We use the \emph{CCEval} dataset, which includes these tuples. We evaluate performance using \emph{Exact Match} (prediction matches ground truth) and \emph{Completion Acceptance Rate (CAR)} (rate of user accepting completions). Detailed descriptions of the \emph{CCEval} dataset and evaluation metrics are provided in sections \ref{sec:multline_dataset} and \ref{sec:metrics}, respectively.

\vspace{-1ex}
\subsection{Need for Curriculum Aware Training}\label{sec:motivation_curiculum}

\subsubsection{Quantitative Analysis:}\label{sec:quan_analysis_car_offline} In our study, we analyzed empirical data from Cody, a popular coding assistant to identify patterns related to code completion acceptance rate (CAR). To collect this data, we implemented the following method:

\begin{enumerate}
    \item At runtime, when a code suggestion query is triggered for a user, we use tree-sitter to parse the code and identify the node type where the completion is requested.

    \item We record whether the completion was accepted by the user and persisted for at least 30 seconds.
\end{enumerate}

\begin{figure}[t]
    \centering
     \vspace{-2ex}
    \includegraphics[width=\linewidth]{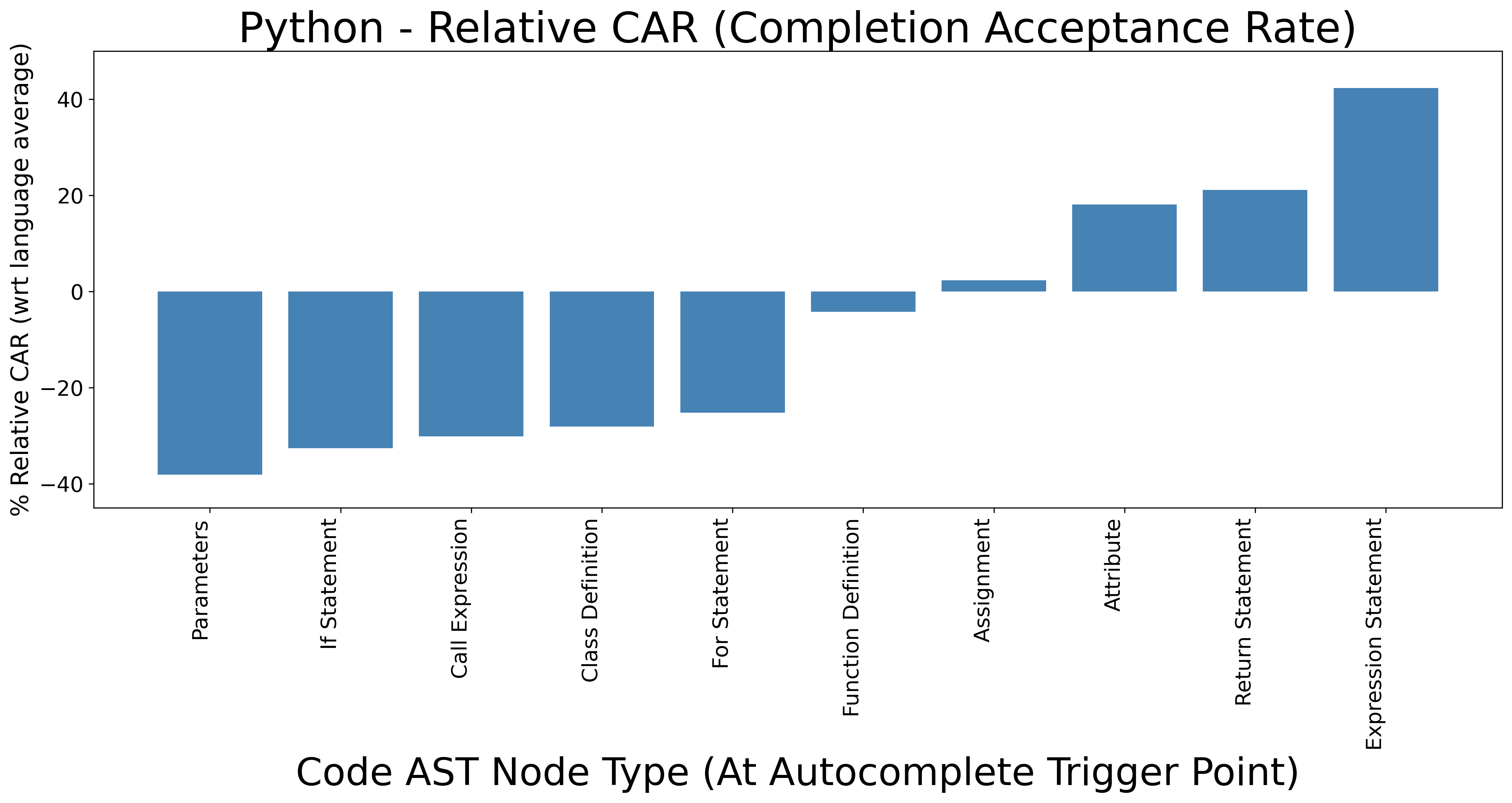}
    \caption{\% Relative CAR at AST nodes: Negative CAR values for node types like \emph{Call Expression} and \emph{Parameters} suggest their higher complexity than average.}
    \label{fig:curriculum_motivation}
\end{figure}

Our findings, presented in Figure \ref{fig:curriculum_motivation}, demonstrate the relative CAR for various node types compared to the language average, with examples of these node types shown in Figure \ref{fig:ast_diagram}. The x-axis displays common node types triggering completions, while the y-axis shows the relative CAR. The analysis reveals that certain node types, such as \emph{If Statement} and \emph{Call Expression}, have significantly lower relative CAR compared to the language average, whereas \emph{Return Statement} have positive CAR than average. We observe similar pattern across multiple programming languages, including JavaScript and TypeScript. We posit that these node types represent more complex patterns, which are more challenging for the model to complete accurately.

\begin{figure}[t]
    \centering
     \vspace{-2ex}
    \includegraphics[width=\linewidth]{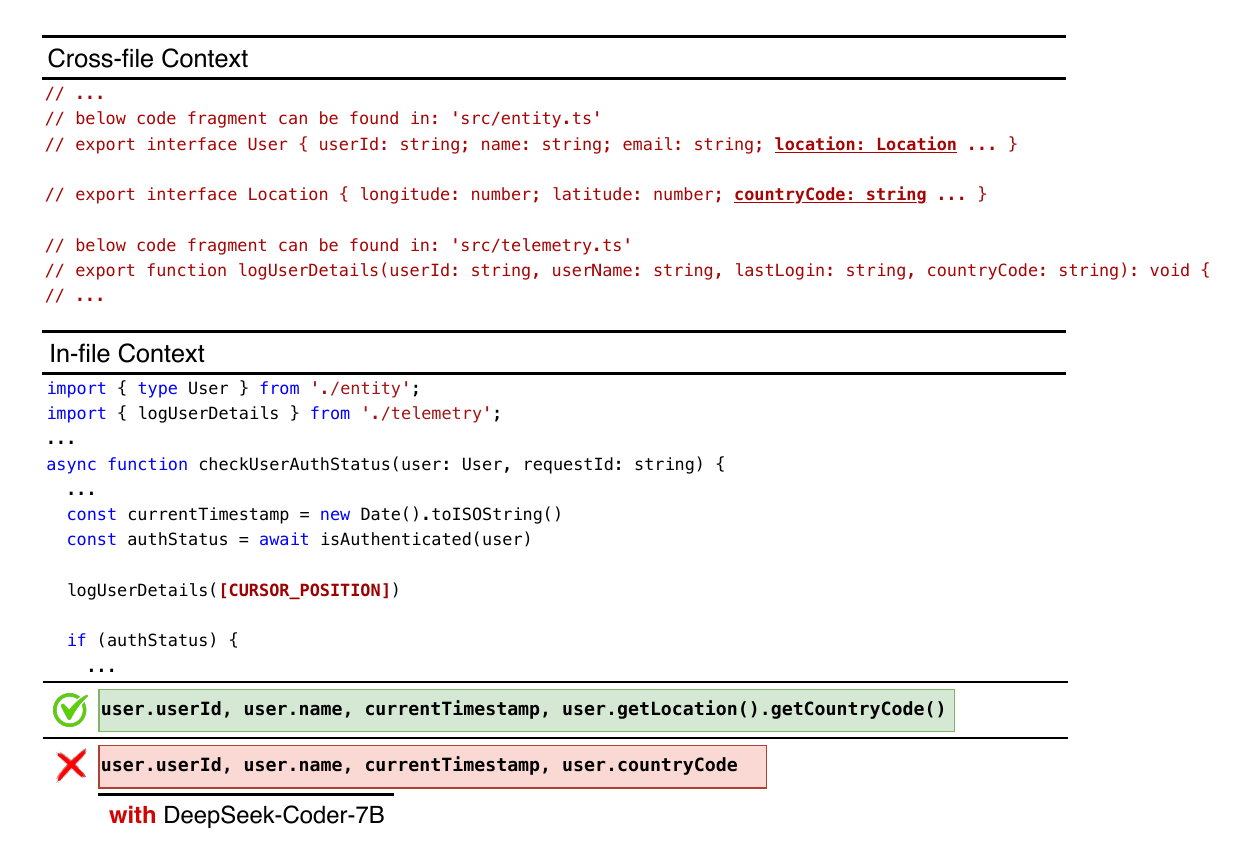}
    \caption{Motivating example for Curriculum  learning: Model fails to predict completion at \emph{Call Expression} node, with nested symbol \emph{countryCode} as an argument.}
    \label{fig:curiculum_motivation_example}
\end{figure}

\subsubsection{Qualitative Analysis:} Complementing our quantitative findings, Our qualitative analysis reveals that different node types present varying levels of complexity for code completion tasks. For instance, \emph{Call Expression} often require context for calling function, its argument, types etc. which are often located in multiple files within the repository, making them more challenging to complete accurately. In contrast, simpler node types like \emph{Return Statement} typically utilize identifiers already present in the function's prefix, making them easier to complete. Figure \ref{fig:curiculum_motivation_example} illustrates one such example of completion at a \emph{Call Expression} node. In this case, the model needs to predict four identifiers, with the last one being particularly challenging. It requires accessing the \emph{countryCode} field from the nested \emph{Location} interface, which is defined in a separate file. The \emph{DeepSeek-Coder-7B} model fails to predict this last identifier correctly, highlighting the difficulty of such completions at this node type.

To enhance the model's performance on node types with lower completion acceptance rates, we propose a Curriculum Aware learning approach, where instead of randomly sampling code from a file as ground truth, we extract patterns similar to these challenging examples. Later, in section \ref{sec:online_exp_results}, we demonstrate that training the model on such \emph{difficult-to-complete} curriculum examples significantly improves CAR overall and across node types that have negative relative CAR.

\vspace{-1em}
\subsection{Need for Context Aware Training}
Fill-in-the-middle (FIM) code language models are typically pre-trained by sequentially appending a set of files and training on that sequence. Early models, such as Starcoder1 ~\cite{li2023starcoder}, utilized this approach by appending random files within the context window during training. However, this method has inherent limitations. The most relevant files are often contained within a single repository. Appending random files from the entire corpus means that the relevant files may not be in the model's context window. Additionally, for small parameter models, the attention mechanism may not effectively focus on the relevant context within these files.

\begin{figure}[t]
    \centering
     \vspace{-2ex}
    \includegraphics[width=\linewidth]{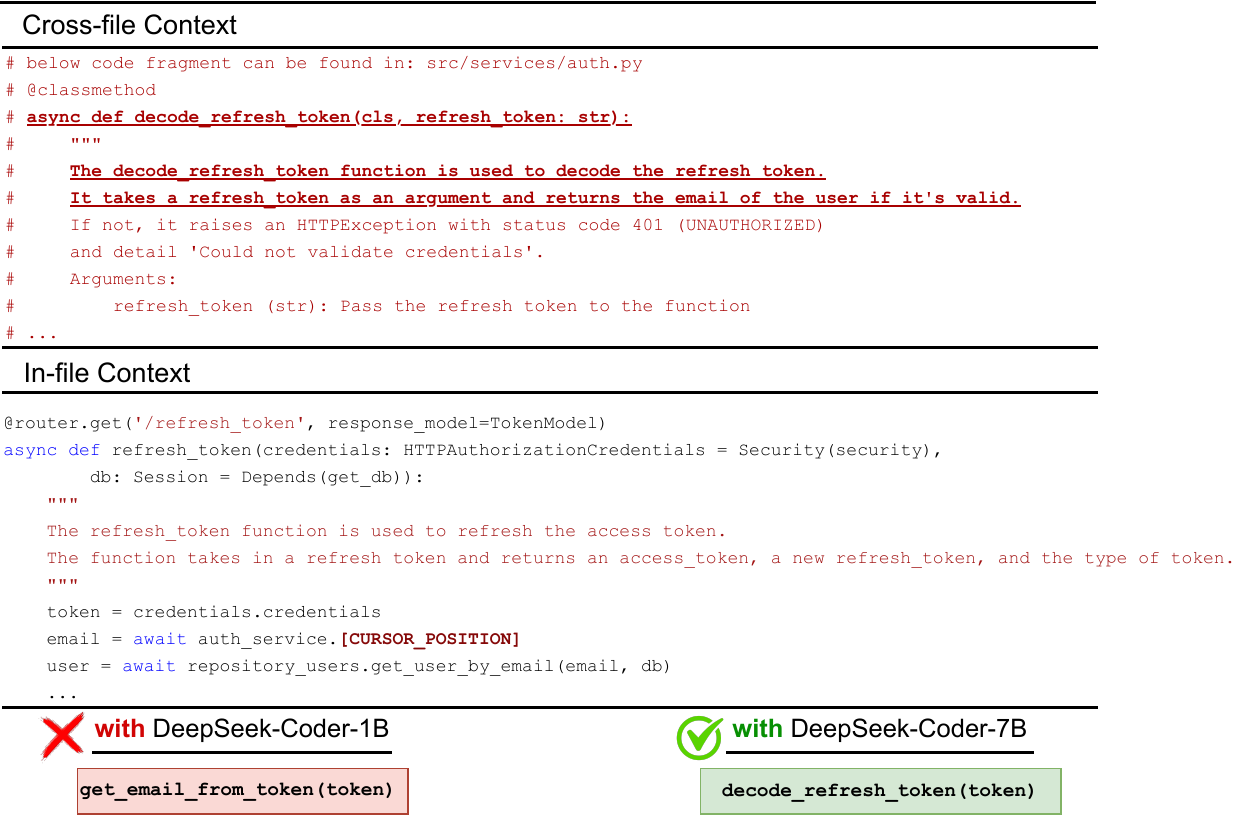}
    \caption{Motivating example for Context learning: Small models often fails to predict completion even with relevant context in prompt.}
    \label{fig:context_motivation_example}
\end{figure}

\begin{figure*}[t]
    \centering
     \vspace{-2ex}
    \includegraphics[width=\linewidth]{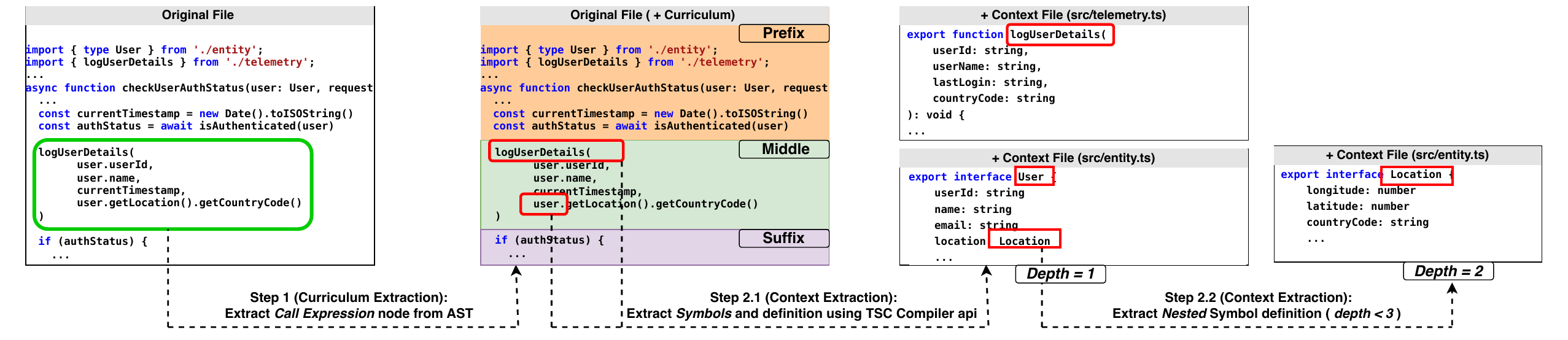}
    \caption{Data generation pipeline illustrating the extraction of curriculum and context examples from source files for \textit{Call Expression} node type, using \textit{tree-sitter} for node extraction and \textit{TSC Compiler API} for precise symbol definitions.}
    \label{fig:data_generation_pipeline}
\end{figure*}

Later models recognized this limitation. For instance, \emph{Starcoder2} ~\cite{lozhkov2024starcoder} was trained using repository-level context, appending files from a single repository together but in a random order. More recent approaches, such as \emph{DeepSeek-Coder} ~\cite{zhu2024deepseek}, further refine this by creating a dependency graph of files based on code imports. Files are then organized and appended according to the sequence derived from the dependency graph rather than in random order.

\subsubsection{Qualitative Analysis:} Smaller parameter size models still face challenges and fail to leverage the context effectively. Figure \ref{fig:context_motivation_example} illustrates an example from the CCEval dataset. We see that the \emph{decode\_refresh\_token} function is the relevant context present in the prompt, with its function definition and documentation reflecting its usage. While we have the relevant context present in the prompt, the \emph{DeepSeek-Coder-1B} incorrectly predicts a non-existing function \emph{get\_email\_from\_token}, whereas \emph{DeepSeek-Coder-7B} was able to correctly predict the function.

\subsubsection{Quantitative Analysis:} We sampled examples from the \emph{CCEval} dataset, where the relevant context is present in the prompt (i.e. Oracle strategy included in the original dataset) and calculated the \emph{Exact Match} metric. Table \ref{tab:context_motivation_analysis} summarizes these results, showing that \emph{DeepSeek-Coder-7B} correctly predicted nearly twice as many examples as \emph{DeepSeek-Coder-1B} across all languages in the dataset. Additionally, we observe that even \emph{DeepSeek-Coder-7B} is unable to get almost half of the predictions correct despite relevant context being present in the prompt.

Given these observations, we hypothesize that training fill-in-the-middle completions by gathering relevant context in the prompt during training could improve the model's ability to utilize context efficiently. Later in section \ref{sec:model_across_param}, we conclude that this strategy significantly increases the performance for smaller models, which are essential for real-time coding assistance but often struggle to leverage context effectively.

\vspace{-1ex}
\section{Method}\label{sec:method}

In this section we describe the approach to extract the Curriculum and Context aware examples. Our data generation pipeline consists of three steps: a) Basic Filtering b) Curriculum Extraction and c) Context Extraction. Our base dataset majorly consists of \emph{the-stack-dedup} ~\cite{kocetkov2022stack3tbpermissively} dataset from BigCode. While the BigCode dataset is sufficiently large, we notice limited number of data points for certain languages such as the React version of TypeScript and JavaScript, MATLAB, Scala, etc., after filtering. To address this, we also scrape open-source GitHub repositories for these specific languages. To filter out low-quality code, we applied same basic filters as used in Starcoder ~\cite{li2023starcoder} and DeepSeek ~\cite{guo2024deepseek} models.

\vspace{-1em}
\subsection{Curriculum Extraction}
As discussed in Section \ref{sec:motivation_curiculum}, certain node types have lower CAR due to their complexity. Traditional FIM code language models use Random Span infilling for pre-training. We hypothesize that including common failure cases in the training dataset can improve model performance in challenging scenarios. To test this, we extract such nodes from code files and label them as ground truth.

Figure \ref{fig:data_generation_pipeline} illustrates our data generation pipeline, with Step 1 showing the curriculum extraction for the \emph{Call Expression} node type. We extract curriculum examples using the following steps:

\begin{enumerate}
    \item \textbf{Code Parsing:} After basic filtering, we parse the code files using tree-sitter to generate Abstract Syntax Tree (AST).

    \item \textbf{Node Extraction:} We extract node types from the code files as specified in Table \ref{tab:curriculum_strategy}. Nodes with text lengths outside the 0.05 to 0.95 quantile range of the training corpus are filtered out.

    \item \textbf{Complexity Sorting:} Node types are ranked by descending complexity, measured by the count of unique symbols (i.e. identifier node type) within its AST. We recursively traverse the AST to extract these symbols. For instance, in Figure \ref{fig:data_generation_pipeline}, the \emph{Call Expression} for \emph{logUserDetails} contains three unique symbols: \emph{logUserDetails}, \emph{user}, and \emph{currentTimeStamp}, giving it a complexity score of 3.
    
    \item \textbf{Sampling and Labeling:} Finally, we sample one node type based on the distribution in Table \ref{tab:curriculum_strategy} as the ground truth. The code before this node becomes the Prefix, and the code after it becomes the Suffix.
    
\end{enumerate}

\subsection{Context Extraction}

After extracting curriculum examples, we obtain relevant context for the ground truth (the "middle") in the Curriculum. Figure \ref{fig:data_generation_pipeline}, Step 2, shows the context extraction process for a \emph{Call Expression} node type. We follow these steps to extract contextual information for the ground truth:

\begin{table}[t!]
\centering
\caption{Exact Match on CCEval: \emph{DeepSeek-Coder-1B} significantly under-performs compared to \emph{DeepSeek-Coder-7B}, despite oracle context in prompt. }
{
\begin{tabular}{lcccc}
\specialrule{.1em}{.05em}{.05em}

Model & Python & Java & Typescript & C\# \\
\specialrule{.1em}{.05em}{.05em}
DeepSeek-Coder-1B & 0.18 & 0.26 & 0.27 & 0.25 \\
DeepSeek-Coder-7B & \textbf{0.35} & \textbf{0.42} & \textbf{0.49} & \textbf{0.51} \\
\hline

\end{tabular}
}
\label{tab:context_motivation_analysis}
\end{table}

\begin{table}[t!]
\centering
 \vspace{-3ex}
\caption{Distribution of AST nodes in the training dataset based on their inference traffic and complexity, specifically focusing on those with high negative CAR.}
{
\begin{tabular}{lc}
\specialrule{.1em}{.05em}{.05em}
Curriculum Strategy & Distribution \\
\hline
Random Span & 35\% \\
Call Expression & 12\% \\
Function Definition (full function) & 12\% \\
Class Definition & 10\% \\
Function Parameters & 8\% \\
Function Definition (with prefix) & 8\% \\
If Statement & 8\% \\
Try-Catch & 4\% \\
Assignment & 3\% \\
\hline
\end{tabular}
}
\label{tab:curriculum_strategy}
\end{table}

\begin{enumerate}
    \item \textbf{Symbol Extraction:} We use tree-sitter to extract symbols from ground truth code snippets. Our goal is to obtain relevant symbol definitions for inclusion in the context.
        
    \item \textbf{Symbol Filtering:} After extracting the symbols, we filter out common English stop words and language-specific keywords.
    
    \item \textbf{Symbol Definition Extraction:}
    We extract symbol definitions from JavaScript and TypeScript files using the TSC compiler API, also highlighted in Figure \ref{fig:data_generation_pipeline}, Step 2.1. For JavaScript-only repositories lacking a \emph{tsconfig.json} file, we create one at the project root with \emph{"allowJs": "true"} to enable API usage. Additionally, We then filter out definitions from common modules like \emph{StdLib}, \emph{Node Modules}. 
    
    \item \textbf{Symbol Graph:} As highlighted in Figure \ref{fig:data_generation_pipeline}, Step 2.2, symbols can have nested definitions, forming a symbol graph, a useful source of information for accurately predicting the ground truth. We extract these nested definitions to a depth of 2, with leaf nodes representing symbols of primitive types. To manage token limits, we use breadth-first search, prioritizing immediate symbols in the graph.
    
    \item \textbf{BM25 Context Retrieval:} For non-TypeScript/JavaScript languages, we employ BM25 for context retrieval. We chunk the repository into syntactic units, preserving whole functions or classes, and create a BM25 index. Large chunks are further divided into smaller AST nodes. To retrieve relevant chunks for the ground truth code, we form a query from filtered symbols and search the BM25 index.
    
    \item \textbf{Preventing Data Leakage:} To prevent data leakage, since a symbol definition or code snippet can sometimes originate from the ground truth itself, we remove such extracted snippets from the context.
    
\end{enumerate}

In summary, the final training dataset is constructed by extracting Curriculum examples from code files and augmenting them with relevant context. Table \ref{tab:curriculum_strategy} summarizes the Curriculum strategy and distribution of node types in the training dataset. Later, in section \ref{sec:applications}, we demonstrate that combining both Curriculum and its associated Context yields greater improvements than deploying either strategy individually, thus showing how these approaches complement each other effectively.

\section{Experimentation} \label{sec:applications}

\begin{table*}[t!]
\centering
\caption{Overall Comparison of \emph{CMFT} with the base models across three datasets. The evaluation metric is \emph{Exact Match} for Single-Line Infilling, \emph{Prefix Match} for CCEval and \emph{Pass@1} for Multi-Line Infilling. The \emph{CMFT} impoves across all the models architectures, parameters and datasets, highlighting its generalizability.}

{
\begin{tabular}{lc|ccc|cccc|c}
\specialrule{.1em}{.05em}{.05em}

\multirow{2}{*}{Model} & \multirow{2}{*}{\#AP} & \multicolumn{3}{c|}{Single-Line Infilling} & \multicolumn{4}{c|}{CCEval} & Multi-Line Infilling \\
& & Python & Java & Javascript & Python & Java & Typescript & C\# & Python \\ 
\hline

DeepSeek-Coder-base & \multirow{2}{*}{1B} & 73.4 & 84.4 & 80.5 & 31.5 & 36.2 & 35.4 & 32.2 & 7.1 \\
+ \textbf{CMFT} & & \textbf{73.9} & \textbf{84.9} & \textbf{81.2} & \textbf{33.3} & \textbf{37.8} & \textbf{36.6} & \textbf{33.6} & \textbf{8.5}  \\
\hline

Starcoder2 & \multirow{2}{*}{3B} & 59.6 & 75.1 & 73.2 & 35.3 & 38.2 & 37.5 & 34.6 & 10.5 \\
+ \textbf{CMFT} & & \textbf{60.6} & \textbf{76.4} & \textbf{73.7} & \textbf{36.1} & \textbf{39.2} & \textbf{38.3} & \textbf{35.5} & \textbf{11.9} \\
\hline

Starcoder2 & \multirow{2}{*}{7B} & 66.1 & 79.3 & 83.6 & 37.4 & 41.9 & 38.8 & 37.6 & 12.7 \\
+ \textbf{CMFT} & & \textbf{66.4} & \textbf{79.8} & \textbf{84.0} & \textbf{37.9} & \textbf{42.2} & \textbf{39.2} & \textbf{38.2} & \textbf{13.4} \\
\hline

DeepSeek-Coder-v2-lite-base & \multirow{2}{*}{2.4B} & 79.8 & \textbf{88.4} & 87.3 & 38.2 & 43.2 & 42.5 & 39.6 & 14.1 \\
+ \textbf{CMFT} & & \textbf{80.3} & 88.3 & \textbf{88.4} & \textbf{38.8} & \textbf{43.9} & \textbf{43.2} & \textbf{40.3} & \textbf{15.5} \\
\hline

DeepSeek-Coder-base & \multirow{2}{*}{7B} & 79.2 & 88.5 & 86.1 & 40.2 & 47.4 & 45.1 & 43.6 & 13.2 \\
+ \textbf{CMFT} & & \textbf{79.8} & \textbf{88.9} & \textbf{86.7} & \textbf{40.8} & \textbf{47.7} & \textbf{45.8} & \textbf{44.1} & \textbf{13.9} \\
\hline

\end{tabular}
}
\label{tab:overall_comparison}
\end{table*}

\subsection{Experimental Setup}

\subsubsection{Datasets: Single-Line FIM} We evaluate single-line fill-in-the-middle completions on two popular public datasets: \emph{Single-Line Infilling Benchmark} and \emph{CrossCodeEval}:

\begin{enumerate}
    \item \textbf{Single-Line Infilling:} This dataset was proposed by Allal et al. ~\cite{allal2023santacoder}, this benchmark masks a single line from the Multipl-E ~\cite{cassano2023multipl} dataset. Models must regenerate the missing line, considering both the preceding and following context.

    \item \textbf{CrossCodeEval (CCEval):} CCEval dataset is derived from real-world repositories in Python, Java, TypeScript, and C\#. It includes code examples where cross-file context is crucial for code completion. The dataset includes prefix, cross-file dependencies, and suffix information at completion points, supporting various code generation tasks, including in-fill scenarios.
    
\end{enumerate}

\subsubsection{Datasets: Multi-Line FIM}\label{sec:multline_dataset}

Existing multi-line evaluation benchmarks such as humaneval-infilling are not derived from real-world code repositories. Instead, they focus on isolated problems such as completing a Fibonacci function, which do not accurately reflect the everyday coding tasks developers encounter. On the other hand, Real-world datasets such as \emph{CCEval} ~\cite{ding2023crosscodeevaldiversemultilingualbenchmark} or \emph{RepoBench} ~\cite{liu2023repobenchbenchmarkingrepositorylevelcode} , typically have single line completion as the ground truth.

To address the gaps in traditional benchmarks and make the evaluation more reflective of real-world scenarios for multi-line fill-in-the-middle, we propose a \emph{Multi-Line Infilling} benchmark derived from the existing \emph{SWE-bench} dataset ~\cite{jimenez2023swe}. SWE-bench is particularly suitable as it provides an execution environment and includes Pull requests with at least one test case transitioning from FAIL to PASS status.




To create a Multi-Line Infilling benchmark from the original SWE-bench dataset, we employ the following process:

\begin{enumerate}
    \item \emph{Filtering Candidate Patches:} We select SWE-bench test instances containing at least one patch set that modifies multiple consecutive lines, which we term candidate patch sets.
    
    \item \emph{Ensuring Test Case Reliability:} We apply all patches except the candidate set to the PR and verify the test case fails. This confirms the test's success depends on the candidate patches. We discard candidates if the test passes without them. The remaining valid candidates become the ground truth patch.

    \item \emph{Isolating the Ground Truth Patch Set:} We apply all patch sets except the ground truth to the PR. The ground truth patch set then represents the code sections the models must complete.

    \item \emph{Model Evaluation:} Models are tasked to predict the consecutive lines needed to complete the code from the ground truth patch set. The generated code replaces the original patch and is evaluated using the SWE-bench framework to determine if it passes the test cases.

\end{enumerate}

In total, we derive 375 problems in the dataset, all of which are in Python, and is publicly available on Hugging Face \footnote{\url{https://huggingface.co/datasets/sourcegraph/code-multi-line-infilling-benchmark}}.


\subsubsection{Models}
We benchmark popular open-source code language models and their fine-tuned variants. We focus on models pre-trained with the FIM objective. Given that latency is a critical factor, we experiment with various model sizes up to 7 billion parameters. To ensure fair comparison, we standardize the context window to 8k tokens across all models. Our evaluation includes the following state-of-the-art open-source models:

\begin{enumerate}
    \item \textbf{Starcoder2} A family of code models with 3B, 7B, and 15B parameters, trained on the Stack2 ~\cite{lozhkov2024starcoder} dataset comprising 4.3 trillion tokens across 619 languages.

    \item \textbf{DeepSeek-coder} A series of code language models trained on 2 trillion tokens (87\% code, 13\% natural language). We evaluate the 1B and 7B variants.

    \item \textbf{Deepseek-coder-v2-lite-base} An MoE code model with 2.8B active parameters and 16B total parameters, trained on an additional 6 trillion tokens beyond DeepSeek-v2.
\end{enumerate}

\begin{table*}[t!]
\centering
\caption{Ablation study on CCEval dataset: Including Curriculum and Context improves Code LM performance, with \emph{CMFT} yielding the best results across various model families and sizes.}
{
\begin{tabular}{lccccccccc}
\specialrule{.1em}{.05em}{.05em}

\multirow{2}{*}{\textbf{Model}} & \multirow{2}{*}{\textbf{\# Active Param}} & \multicolumn{2}{c}{\textbf{Python}} & \multicolumn{2}{c}{\textbf{Java}} & \multicolumn{2}{c}{\textbf{Typescript}} & \multicolumn{2}{c}{\textbf{C\#}} \\

& & PM & ES & PM & ES & PM & ES & PM & ES \\
\hline

DeepSeek-Coder-base &  1B &  31.5\% &  51.8\% &  36.2\% &  53.5\% &  35.4\% &  52.5\% &  32.2\% &  51.2\% \\

 + \textbf{RSFT} &  &  31.3\% &  51.9\% &  36.8\% &  53.4\% &  35.6\% &  52.3\% &  32.8\% &  51.5\% \\
 + \textbf{CUFT} &  &  31.8\% &  52.9\% &  37.1\% &  54.2\% &  35.8\% &  53.6\% &  32.9\% &  51.9\% \\ 
 + \textbf{COFT} &  &  32.3\% &  52.4\% &  37.4\% &  54.5\% &  36.1\% &  53.7\% &  32.9\% &  52.5\% \\ 
 + \textbf{CMFT} &  &  \textbf{33.3}\% &  \textbf{53.4}\% &  \textbf{37.8}\% &  \textbf{54.5}\% &  \textbf{36.6}\% &  \textbf{54.0}\% &  \textbf{33.6}\% &  \textbf{52.6}\% \\ 
\hline

%


 Starcoder2 &  7B &  37.4\% &  59.6\% &  41.9\% &  62.5\% &  38.8\% &  52.0\% &  37.6\% &  58.3\% \\ 

 + \textbf{RSFT} &  &  37.6\% &  59.6\% &  41.8\% &  62.3\% &  38.7\% &  52.2\% &  37.5\% &  58.4\% \\ 
 + \textbf{CUFT} &  &  37.8\% &  59.8\% &  42.1\% &  63.0\% &  39.0\% &  52.4\% &  37.8\% &  58.7\% \\ 
 + \textbf{COFT} &  &  \textbf{38.0}\% &  \textbf{60.1}\% &  42.1\% &  62.7\% &  39.2\% &  52.4\% &  38.0\% &  58.8\% \\ 
 + \textbf{CMFT} &  &  37.9\% &  60.0\% &  \textbf{42.4}\% &  \textbf{63.0}\% &  \textbf{39.2}\% &  \textbf{52.6}\% &  \textbf{38.2}\% &  \textbf{59.1}\% \\ 
\hline

 DeepSeek-Coder-v2-lite-base &  2.4B &  38.2\% &  60.1\% &  43.2\% &  61.5\% &  42.5\% &  59.5\% &  39.6\% &  60.2\% \\ 

 + \textbf{RSFT} &  &  38.4\% &  60.3\% &  43.2\% &  61.4\% &  42.7\% &  59.8\% &  39.6\% &  60.3\% \\ 
 + \textbf{CUFT} &  &  38.7\% &  60.6\% &  43.7\% &  61.7\% &  43.0\% &  59.8\% &  39.8\% &  60.4\% \\ 
 + \textbf{COFT} &  &  38.7\% &  60.7\% &  43.7\% &  \textbf{62.0}\% &  42.8\% &  59.8\% &  40.1\% &  60.6\% \\ 
 + \textbf{CMFT} &  &  \textbf{38.8}\% &  \textbf{60.7}\% &  \textbf{43.9}\% &  \textbf{62.0}\% &  \textbf{43.2}\% &  \textbf{60.1}\% &  \textbf{40.3}\% &  \textbf{60.7}\% \\ 
\hline

\hline

\end{tabular}
}
\label{tab:cceval_ablation_offline_result}
\end{table*}

\vspace{-1em}
\subsubsection{Fine-tuning details} 

To showcase the benefits of incorporating curriculum and context into training regime (Section \ref{sec:method}), we fine-tuned the model highlighted in the previous section on the generated dataset from section \ref{sec:method}. We investigated various model architectures and sizes to evaluate the performance across different dimensions. For each base model, we fine-tuned four variants:

\begin{enumerate}
    \item Random Span Fine-Tuning (\textbf{RSFT}): Models are fine-tuned using random code spans as ground truth. This method serves to evaluate potential differences arising from fine-tuning of the models without Curriculum/Context.
    
    \item Curriculum-Aware Fine-Tuning (\textbf{CUFT}): This technique involves fine-tuning models exclusively on curriculum-based examples, focusing on structured learning progression.
    
    \item Context-Aware Fine-Tuning (\textbf{COFT}): Models are fine-tuned exclusively on context examples. We select the ground truth for each file using Random Span. 
    
    \item Combined Fine-Tuning (\textbf{CMFT}): This method integrates both curriculum and context-aware approaches, fine-tuning models on a combination of curriculum-based and context-rich examples.
    
\end{enumerate}

For memory efficiency, We applied LoRA ~\cite{hu2021lora} with $r=8$ and $\alpha=16$. We employed a batch size of 16, learning rate of $3e^{-4}$, and maximum 1 epoch, with a 16,000 token limit per model. Our training pipeline included approximately 200k data points. As the base models were pre-trained on FIM, we used a fixed $fim\_rate=0.5$, following other models \cite{bavarian2022efficient}, without further experimentation due to computational constraints. We employed the PSM (Prefix-Suffix-Middle) mode of the FIM methodology for fine-tuning, as it has shown better empirical performance \cite{bavarian2022efficient}. We provider Prompts in the same format as the initial pre-training for the model.


\subsubsection{Evaluation Metrics}\label{sec:metrics}

For offline experiments, we evaluate single-line and multi-line completions differently. Single-line completions use Exact Match (EM) and Edit Similarity (ES), with Prefix Match (PM) for the CCEval dataset. Multi-line completions are evaluated using Pass@1.

For online experiments, we evaluate completions using user feedback metrics: Completion Acceptance Rate (CAR) and Completion Persistence Rate (CPR).

\begin{enumerate}
    \item \emph{Prefix Match (PM)}: We check if the ground truth code forms the prefix of the generated output. This aligns with real-world code completion assistants that display the next few lines to users, instead of whole generated output.
    
    \item \emph{Completion Acceptance Rate (CAR)}: CAR is defined as the ratio of number of accepted completions to number of suggestions shown to users for more than 750ms.

    \item \emph{Completion Persistence Rate (CPR)}: CPR is measured by normalizing the edit distance between the accepted completion and surrounding code after set time intervals (such as 30, 120, and 300 seconds). We consider a completion persisted if the edit distance is below 0.33.
    
\end{enumerate}

\subsection{Offline Evaluation}\label{sec:offline_eval}

\subsubsection{Overall Comparison}
Table \ref{tab:overall_comparison} summarizes the performance of the \textbf{CMFT} method across three datasets. We use Exact Match for Single-Line Infilling, Prefix Match for CCEval, and Pass@1 for Multi-line Infilling. For single-line datasets, we only consider the first line of the generated code and disregard the rest.

Our results indicate that among base models, DeepSeek-Coder-v2-lite-base demonstrates superior performance on Single-Line and Multi-Line Infilling benchmarks, while DeepSeek-Coder-Base (7B parameters) performs best on the CCEval dataset. Fine-tuning with \textbf{CMFT} consistently enhances performance across all base models, suggesting effective context utilization. The observed improvements across all three datasets indicate the method's generalizability to various models and dataset types.

\begin{figure}[t]
    \centering
     \vspace{-2ex}
    \includegraphics[width=1.0\linewidth]{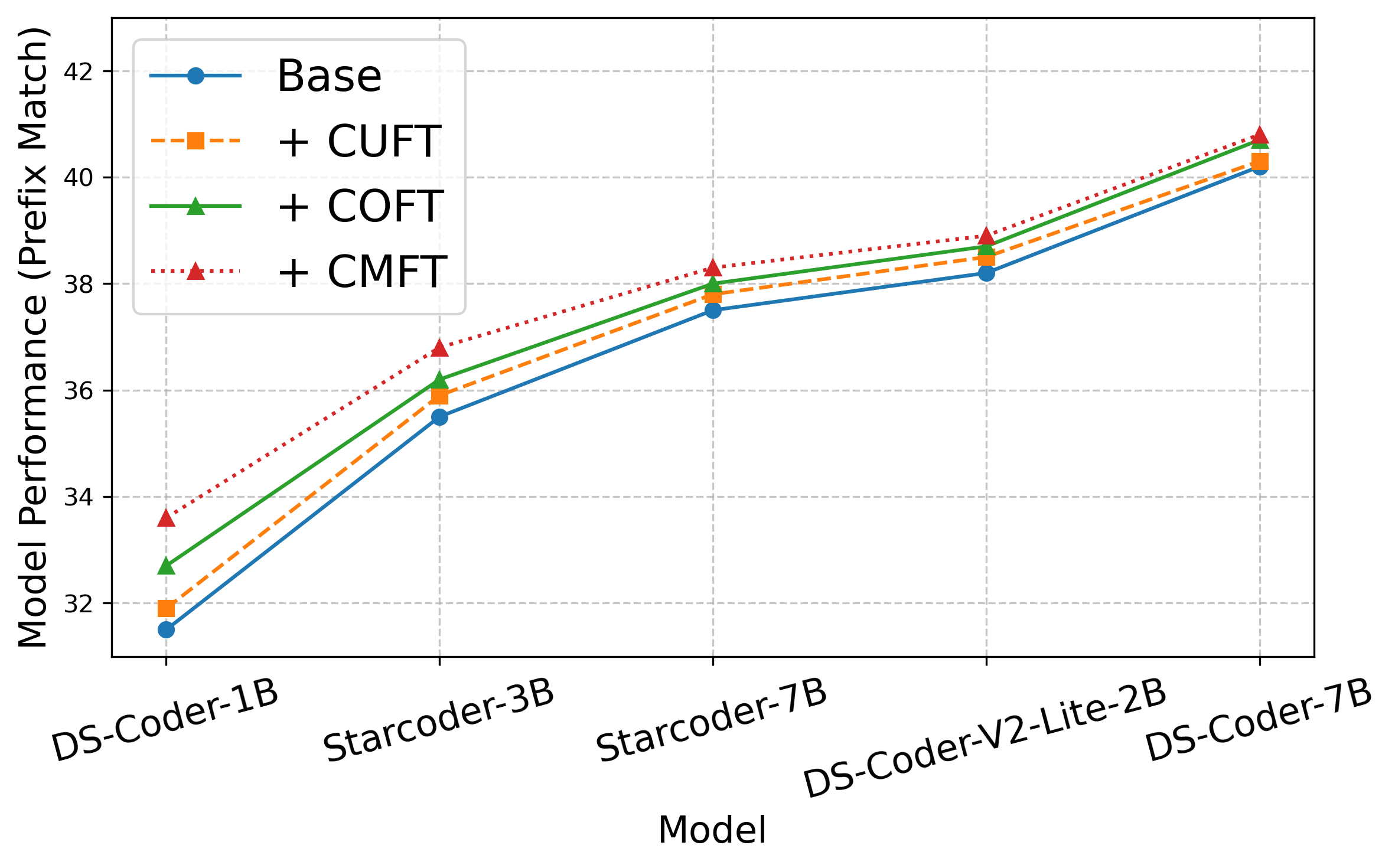}
    \caption{Prefix Match on CCEval dataset for Python: Improvements observed across all model sizes, with smaller parameter models showing a larger improvement delta}
    \label{fig:cceval_typescript_prefix_match}
\end{figure}

\subsubsection{Ablation Study}
\textbf{CMFT} comprises Curriculum Fine-Tuning (\textbf{CUFT}) and Context Fine-Tuning (\textbf{COFT}). While base models are pre-trained with random code spans during FIM, continued fine-tuning can affect performance. We include \textbf{RSFT} in our ablation study to assess this impact. As CCEval is derived from real repositories and requires cross-file context, we conduct ablation studies on it to quantify each module's performance. We select DeepSeek-Coder-1B, DeepSeek-Coder-v2, and Starcoder2-7B for our study, as they represent diverse model families, parameter sizes, and architectures, with DeepSeek-Coder-v2 being a MoE model.

Table \ref{tab:cceval_ablation_offline_result} presents the ablation study results. \textbf{RSFT} shows performance similar to the base model, indicating minimal gain from fine-tuning on random span data, aligning with Bavarian \emph{et al.} ~\cite{bavarian2022efficient} findings. Both \textbf{CUFT} and \textbf{COFT} outperform the base models and \textbf{RSFT}, suggesting effective learning from challenging examples and improved context utilization.

Furthermore, \textbf{COFT} generally outperforms \textbf{CUFT}, suggesting that incorporating context during training yields more significant improvements than curriculum-based fine-tuning. \textbf{CMFT}, which combines both context and curriculum fine-tuning, shows the most substantial gains in nearly all cases, indicating that both approaches complement each other. This aligns with our data extraction methodology, where we first extract examples for completion and then extract relevant context for those examples.

\vspace{-1ex}
\begin{table}[t!]
\centering
\caption{\emph{Pass@1} and \emph{Latency} on Multi-Line Infilling benchmark: Although \emph{CMFT} underperforms compared to Claude and GPT but balances both latency and accuracy.}
{
\begin{tabular}{lcc}
\specialrule{.1em}{.05em}{.05em}

Model & \emph{Pass@1} & Latency (sec) \\
\specialrule{.1em}{.05em}{.05em}

Claude-3.5-sonnet & \textbf{23.4}\% & 2.85 \\
Gpt-4o & 19.4\% & 1.67 \\
Mixtral-8x22b & 12.8\% & 1.38 \\
\hline
DeepSeek-Coder-v2-lite-base & 14.1\% & 0.84 \\

+ \textbf{CUFT} & 14.7\% & 0.86 \\
+ \textbf{COFT} & 15.1\% & \textbf{0.83} \\
+ \textbf{CMFT} & 15.5\% & 0.85 \\
\hline
DeepSeek-Coder-7b & 13.2\% & 1.14 \\

+ \textbf{CUFT} & 13.5\% & 1.21 \\
+ \textbf{COFT} & 13.9\% & 1.19 \\
+ \textbf{CMFT} & 13.9\% & 1.16 \\
\hline

\end{tabular}
}
\label{tab:multiline_infill_eval}
\end{table}

\subsubsection{Model Performance Across Model Size}\label{sec:model_across_param}

Figure \ref{fig:cceval_typescript_prefix_match} illustrates the Prefix Match performance of various models on the CCEval dataset. DeepSeek-Coder-v2, a Mixture of Experts (MoE) model, utilizes 2.4B active parameters during inference from a total of 16B parameters. Our results demonstrate consistent improvements across all base models when fine-tuned on either \textbf{CUFT} or \textbf{COFT} datasets. Notably, smaller models demonstrate more substantial relative improvements, with these benefits diminishing as model size increases. For instance, the \textbf{CMFT} strategy yielded a $6.25\%$ performance gain with \emph{DS-Coder-1B}, but only a $1.25\%$ improvement with the larger \emph{DS-Coder-7B} model. This finding suggests that for code completion tasks, where speed is crucial, we can significantly enhance smaller models' performance without compromising latency. However, it's important to note that fine-tuning alone does not enable smaller models to match the performance of larger ones.
Additionally, we observe that DeepSeek-v2, despite being an MoE model, demonstrates comparable performance to Starcoder-7B with lower inference latency, highlighting the potential benefits of MoE architecture in this context.

\vspace{-1ex}
\subsubsection{Multi-Line Completions: Latency-Performance Trade-off}
We evaluate larger models for multi-line completions, where higher accuracy is expected with increased latency budget. For chat models without native infilling support, we add in-context learning examples. Table \ref{tab:multiline_infill_eval} presents Pass@1 scores and latency for various models, measured by generating 256 tokens with 1500 tokens in context and $temperature=0.2$. We see that adding context and curriculum examples during training enhanced performance without increasing latency. Although, Fine-tuned models show significantly lower latency compared to proprietary models like GPT-4 and Claude-3.5-sonnet, they fall within acceptable constraints for multi-line generation.


\begin{table}[t!]
\centering
\caption{Online A/B: \% change in CAR and Persistence Rate compared to Control. All results are statistically significant (2 tailed t-test at p<0.05 after Bonferroni correction) except for those marked by$^+$.}
{
\begin{tabular}{llcccc}
\specialrule{.1em}{.05em}{.05em}

\multirow{2}{*}{Line Group} & \multirow{2}{*}{Variant} & \multirow{2}{*}{CAR} & \multicolumn{3}{c}{Persistence Rate (sec)} \\
& & & 30 & 120 & 300 \\
\hline

\multirow{3}{*}{Single-Line} & \textbf{+CUFT} & 2.57 & 1.64 & 1.46 & 1.29 \\
 & \textbf{+COFT} & 3.24 & 1.94 & 1.70 & 1.34 \\
 & \textbf{+CMFT} & \textbf{3.97} & \textbf{2.15} & \textbf{1.73} & \textbf{1.64} \\
\hline
\multirow{3}{*}{2-5 Lines} & \textbf{+CUFT} &  3.15 & 1.73 & 1.66 & 1.29 \\
 & \textbf{+COFT} &  3.74 & 2.14 & 1.97 & 1.64 \\
 & \textbf{+CMFT} & \textbf{4.24} & \textbf{2.53} & \textbf{2.1} & \textbf{1.84} \\
\hline
\multirow{3}{*}{5+ Lines} & \textbf{+CUFT} & 1.24 & 0.52$^+$ & 0.31$^+$ & 0.24$^+$ \\
 & \textbf{+COFT} & 2.15 & \textbf{1.32} & 0.72 & \textbf{0.49}$^+$ \\
 & \textbf{+CMFT} & \textbf{2.92} & 1.27 & \textbf{0.95} & 0.41$^+$ \\
\hline

\end{tabular}
}
\label{tab:ab_online_results}
\end{table}

\vspace{-1em}
\subsection{Online Evaluation}
\subsubsection{Experiment Setup:} We conducted an A/B test using DeepSeek-Coder-1.6B as the base model, comparing three fine-tuned variants: a) \textbf{CUFT}, b) \textbf{COFT}, and c) \textbf{CMFT}. The experiment ran for two weeks, with approximately 350,000 suggestions per variant. Each model was deployed on four NVIDIA H100 80GB GPUs with optimizations including Flash Attention 2, n-gram speculation, and prompt caching. We performed load testing to determine optimal GPU and replica configurations. The completion request is typically triggered with each character typed by the user, with additional caching to reuse already calculated suggestions.

\subsubsection{Experiment Results:}\label{sec:online_exp_results} 
\begin{figure}[t]
    \centering
     \vspace{-2ex}
    \includegraphics[width=1.0\linewidth]{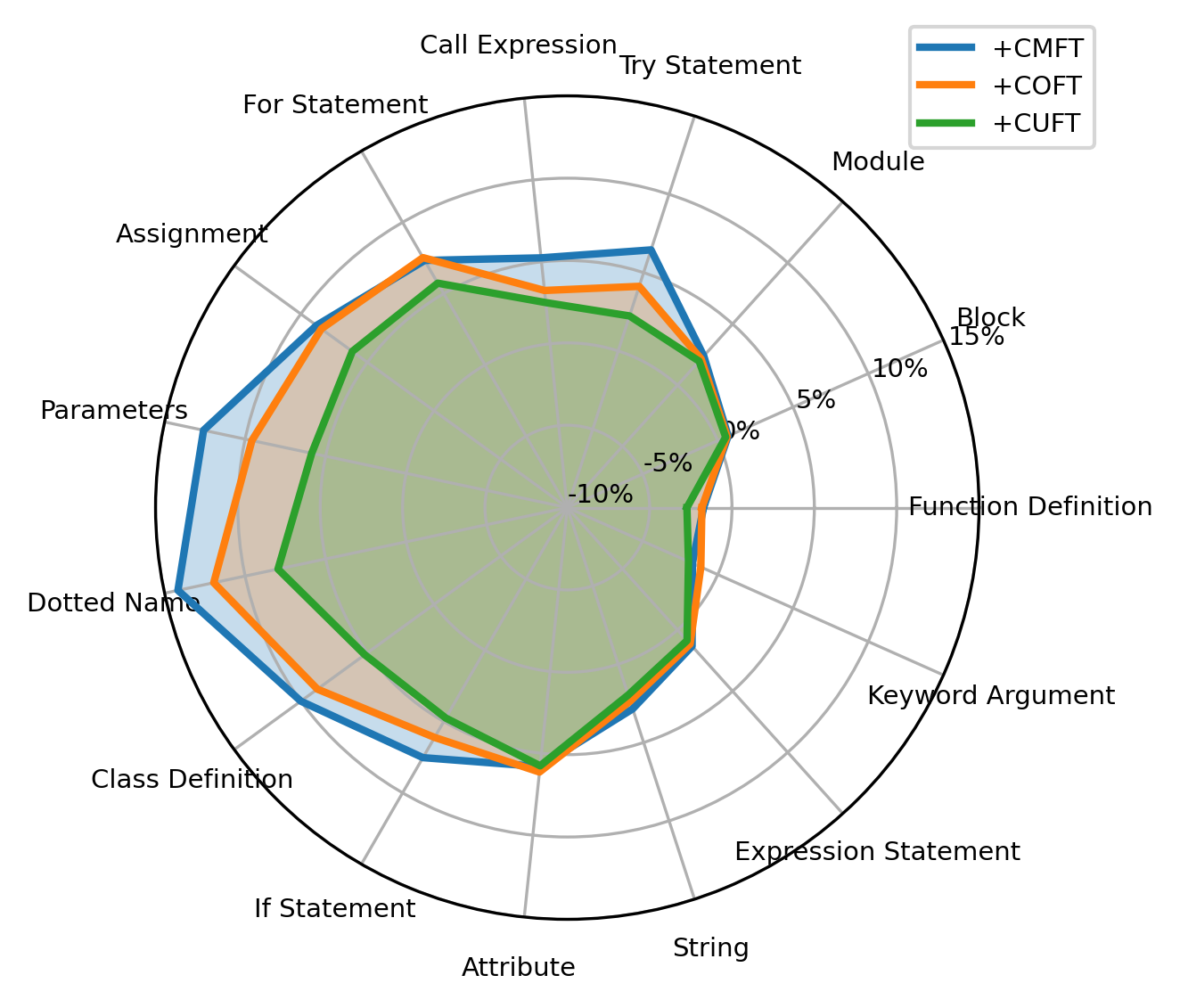}
    \caption{Relative CAR (\%) at AST nodes: Fine-tuned models show significant CAR improvements at complex nodes (referenced in \ref{fig:curriculum_motivation}) compared to the control.}
    \label{fig:car_across_node_types}
\end{figure}

Table \ref{tab:ab_online_results} shows the percentage change in CAR and CPR (Persistence rate) for each fine-tuned model compared to the base DeepSeek-Coder-1.6B model. Results are categorized by Line Group, with 86\% of completions for single lines, 9\% for 2-5 lines, and 5\% for 5+ lines across all variants. The p75 latency metric was maintained for fine-tuned variants.
All variants demonstrated statistically significant improvements in CAR and Persistence Rate. The \emph{CMFT} model performed best, followed by \emph{COFT} and \emph{CUFT}, aligning with offline findings. The most substantial improvements occurred in the 2-5 line group, suggesting enhanced multi-line completion capabilities. Across languages, the \emph{CMFT} demonstrate superior performance on TypeScript, TypescriptReact, JavascriptReact, and JavaScript compared to other languages. This improvement can be attributed to higher-quality data generation, where precise symbol definitions are extracted instead of relying on BM25 for these languages. We revisit the analysis from Section \ref{sec:quan_analysis_car_offline}, comparing the relative CAR across node types for the three variants against the base DeepSeek-Coder-1.6B model (Figure \ref{fig:car_across_node_types}). We see significant improvements on node types, previously under-performing compared to the language average. This suggests that the Curriculum and Context-aware training regime enhances the model's ability to leverage context and comprehend more challenging examples.

Overall, all fine-tuning strategies significantly improved Code Acceptance Rate (CAR) and Persistence Rate. The \textbf{CMFT} demonstrated superior performance across programming languages and AST node types, enhancing contextual understanding and handling complex examples. Furthermore, our analysis revealed a strong positive pearson correlation ($r = 0.6$, $p < 0.05$) between our primary offline metric (Prefix Match) and online metric (CAR), indicating strong alignment between offline and online performance indicators.

\vspace{-2ex}
\section{Discussion \& Future Work} \label{sec:conclusion}

In conclusion, we present a novel approach to enhance Fill-in-the-Middle (FIM) code completion models by incorporating context and curriculum examples in the training process. We demonstrates significant improvements in completion accuracy across various model sizes, with smaller models showing more pronounced gains and validate the effectiveness of this approach is through extensive offline and online A/B evaluations.
We envision future extensions to formalize our work on quantifying code complexity, to get more insights into LLM failure cases. We are also keen on investigating the interpretability of ML models, particularly the layer-level changes during fine-tuning. This could help us understand why improvements diminish  as model size increases and how to overcome this limitation. Lastly, We aim to delve deeper into more nuanced metrics for online A/B testing to gain a more comprehensive understanding of our model's real-world impact.

\section*{Ethical Considerations}
Our research aims to enhance code completion capabilities, potentially improving developer productivity across various programming languages. However, we acknowledge that the introduction of new code datasets and advanced language models carries inherent risks. While our multi-line infilling evaluation benchmark is carefully curated, there remains a possibility that malicious actors could exploit the system to generate harmful code. To mitigate this risk, we have implemented stringent safeguards in our data collection and model training processes, including thorough code review and sanitization procedures.

Furthermore, we recognize that our approach relies on proxy metrics such as Completion Acceptance Rate (CAR) and Completion Persistence Rate (CPR) to measure the effectiveness of code suggestions. While these metrics provide valuable insights, we acknowledge that an overemphasis on such signals could potentially lead to biased or suboptimal code suggestions. To address this concern, we have incorporated diverse evaluation methods, including offline benchmarks and real-world A/B testing, to ensure a more comprehensive assessment of our models' performance and potential impact on developer workflows.

\bibliographystyle{ACM-Reference-Format}
\bibliography{references}

\end{document}